\documentclass{elsart}
\usepackage{graphicx,amssymb}
\journal{Physica A}

\begin{document}

\begin{frontmatter}

\title{Schematic models for fragmentation of brittle solids
       in one and two dimensions}

\author{F.P.M. dos Santos},
\author{R. Donangelo}, and
\author{S.R. Souza}
\address{Instituto de F\'\i sica,
Universidade Federal do Rio de Janeiro,
Caixa Postal 68528, 21941-972 Rio de Janeiro - RJ, Brazil}

\begin{abstract}
Stochastic models for the development of cracks in 1 and 2 dimensional
objects are presented.
In one dimension, we focus on particular scenarios
for interacting and non-interacting fragments during the breakup
process.
For two dimensional objects, we consider only non-interacting fragments,
but analyze isotropic and anisotropic development of fissures.
Analytical results are given for many observables.
Power-law size distributions are predicted for some of the
fragmentation pictures considered.
\end{abstract}

\end{frontmatter}

\section{Introduction}
The breakup of a system into many pieces, {\it i.e.} fragmentation, is
a subject of intensive investigation in different areas of science and
engineering \cite{dynFragReview1999,Bondorf,Gross,fragMolecules}.
The interest in this phenonemon ranges from studies of the observed
size distributions of atomic nuclei \cite{Bondorf,Gross} and
chains of molecules \cite{fragMolecules}
to asteroids \cite{asteroidFrag2000},
ice floes \cite{ice12004,ice22005,ice32005},
brittle solids
\cite{fragBrittle1,fragBrittle2,fragBrittle3,fragBrittle4,fragBrittleFluid1},
thin glass plates\cite{fragGlass2002},
eggshells \cite{egg1,egg2},
frozen potatoes \cite{FragBohr1993}, fluids \cite{fragBrittleFluid1},
drops in general \cite{fragPauloMurilo1,fragPauloMurilo2}, etc.
The size distributions observed in the breakup of these systems exhibit,
in general, a power-law behavior, suggesting, in some cases, at least, that the
process is related to critical phenomena.

This peculiarity has led to the development of many schematic models which,
to a large extent, disregard microscopic details of the fragmenting
macroscopic objects.
The fact that the size distributions seem to be fairly insensitive to the
constituent elements of the fragmenting body \cite{FragBohr1993}
gives strong support to this approach.
Therefore, different treatments, based on quite general assumptions,
have been proposed to explain the characteristics of the fragmentation of
brittle solids.
These approximations include, for instance, mean-field treatments
\cite{meanField1,meanField2,meanField3,meanField4}, fractal analysis
\cite{fragGlass2002,fractalKim1995}, dynamical models of granular solids
\cite{granularSolids1}, random forces stopping models \cite{fragRNDForces1},
and schematic branching-merging models \cite{unifrag,dynFrag2} (for a
recente review see \cite{dynFragReview1999} and references therein).

In this work, we present schematic models for the breakup of rods and
flat brittle objects.
For the latter case, our models are intended to describe the fragmentation
of objects which suffer a strong impact on one of its sides, in
contrast to those proposed in
\cite{meanField1,meanField2,meanField3,meanField4,unifrag}, in which
the stress is uniformly distributed inside the body.
Effects associated with anisotropy are also investigated.
In the one-dimensional case, we compare the properties of the fragments
produced when the interaction between them can be neglected, {\it i.e.}
when they are ejected from the parent body, with the case in which
they continue interacting during the breakup process.

In sect.\ \ref{sect:models} we present the models and make
analytical predictions related to the size distributions,
whereas in sect.\ \ref{sect:results} we discuss and interpret the results.
Conclusions are drawn in sect.\ \ref{sect:conclusions}.

\section{The models}
\label{sect:models}
The main characteristic of the models presented below is that they can
be viewed as stochastic processes, in which strict mass conservation is
imposed in each event.
We present below their detailed formulation.

\subsection{1-dimensional models}
\label{sec:mod1d}
Objects whose lengths are large compared with their cross-sections
are represented by one-dimensional segments of a line.
Thus, in this subsection we consider fractures on a line of unitary length.
The number of fractures $N$ is related to the violence of the impact on
the object.
More specifically, $N$ points $\{x_i\}$, $i=1\dots N$, are associated
with the fractures and two adjacents points delimit a fragment, so that
$N+1$ fragments are formed at the end of the process.

We concentrate on two different fragmentation models, which aim at describing
distinct breakup scenarios:
\begin{description}
\item{\bf i-} Non-Interacting Fragments (NIF1), where fracture points
are sequentially generated and the $(i+1)$-{\it th} crack can only appear at
the right hand side of the $i$-{\it th} fissure.
It is numerically implemented by selecting the first random point $x_1$ 
uniformly in the interval $(0$,$1)$.
Then, the next crack point $x_2$ is uniformly chosen in the interval
$(x_1$,$1)$, and so on.
This corresponds to a simplified picture for the breakup of an object
when the impact zone is concentrated close to its left edge, similar
to the (almost) perpendicular fall of a rod.
In this case, the dynamics cannot produce further fractures in fragments that
have already been released from the parent fragment.
\item{\bf ii-} Interacting Fragments (IF1), in which the broken pieces exert
stress on the others, leading to further cracks.
For simplicity, at the $i$-{\it th} step, one of the $i+1$ fragments is
chosen with equal probability and a fissure point is sampled uniformly
inside the selected fragment.
\end{description}

The analytical construction of the NIF1 model can be achieved through the
following considerations:
\begin{description}
\item{\bf 1-} When the $i$-{\it th} crack point is made, there exists an
equivalence between the fragment just formed and the remaining of the
parent body.
\item{\bf 2-} Let us assume that, after the $i$-{\it th} fracture,
the parent body has length $l$.
Then, as the $(i+1)$-{\it th} fissure is produced, the probability of it
having length $\chi$ must be inversely proportional to its parent's length
$l$.
\end{description}

\noindent
Upon denoting by $P_{i}(\chi)$ the probability density of creating a fragment
of size $0<\chi<1$ at the $i$-{\it th} crack, one may then write:

\begin{equation}
P_{i+1}(\chi)=\int_\chi^1\,\frac{1}{l}P_i(l)\,dl\;.
\label{eq:recnif1}
\end{equation}

\noindent
By noticing that $P_1(\chi)=1$, this recurrence relation can be iterated,
leading to:

\begin{equation}
P_i(\chi)=\frac{\left[-\log(\chi)\right]^{i-1}}{(i-1)!}\;,
\label{eq:probnif1}
\end{equation}

\noindent
so that the probability density of observing a fragment of size $\chi$,
produced at any stage of the fragmentation process is:

\begin{equation}
P(\chi)=\frac{1}{N}\sum_{i=1}^NP_i(\chi)=\frac{1}{N}\sum_{i=1}^N
\frac{\left[-\log(\chi)\right]^{i-1}}{(i-1)!}\;.
\label{eq:spronif1}
\end{equation}

\noindent
For $N$ large, the above expression can then be approximated by:

\begin{equation}
P(\chi)\approx\frac{1}{N}\exp[-\log(\chi)]=\frac{1}{N}\chi^{-1}\;.
\label{eq:sumprobnif1}
\end{equation}

It is worth mentioning that normalization is apparently lost in this
approximation, since the integral of $1/\chi$ from 0 to 1 diverges.
However, the above expression is strictly valid only for
$N\rightarrow \infty$ and,
therefore, the $1/N$ factor would cancel out the divergency.
Thus, this expression should be considered as an approximation for
finite (but large) $N$.
Nevertheless, the results shown in the next section reveal that this
formula is fairly accurate for modest values of $N$.

One should notice that these results are qualitative different from those
corresponding to uniform fragmentation \cite{unifrag}, which should take
place when the object suffers a violent impact equally distributed over its
length.
In this case, the probability density of finding a fragment of size
$\chi$, after $N$ fractures is \cite{unifrag}:

\begin{equation}
P_N(\chi)=N(1-\chi)^{N-1}\;.
\label{eq:probhom}
\end{equation}

The analysis of the IF1 model is more involved.
Although we could not obtain closed expressions, we found useful
recurrence relations which allow one to compute the properties of the
model very easily.
First, one should notice that when a fragment is cracked, the probability
distribution for the size of the generated fragment is equal to that of the
remnant.
Then, the probability density of having a fragment of size $\chi$ after the
first crack is made, $\Psi_1(\chi)$, can be expressed as:

\begin{equation}
\Psi_1(\chi)=\frac{1}{2}[P_1(\chi)+P_1(\chi)]\;,
\label{eq:p1if1}
\end{equation}

\noindent
where $P_i(\chi)$ is given by Eq.\ (\ref{eq:probnif1}).
The normalization factor $1/2$ is associated with the number of fragments
formed after the first crack of the system.

When the second fracture is made, there are two contributions which should be
taken into account:

\begin{description}
\item{\bf a)} a fragment of size $\chi$ can be produced by the crack of
a fragment of size $l$, $\chi < l < 1$, or
\item{\bf b)} the cracking of a selected fragment may lead to a piece of any
size, but the unbroken fragment has size $\chi$.
\end{description}

\noindent
Since (a) can also lead to a fragment of size $1-\chi$ and a remaining part of
size $\chi$, which are statistically equivalent, one may write:

\begin{equation}
\Psi_2(\chi) = \frac{1}{3}\left[2\int_\chi^1\,\frac{1}{l}\Psi_1(l)\,dl
+(2-1)\Psi_1(\chi)\right]\;,
\label{eq:pp2_1}
\end{equation}

\noindent
where the first term accounts for the contribution of (a) and the second one
is associated with (b).
The factor $(2-1)$ accounts for the degeneracy of the remaining
fragments in (b), recalling their statistical equivalence.
The above expression can be further expanded through the use of
Eqs.\ (\ref{eq:recnif1}) and (\ref{eq:p1if1}), and one obtains:

\begin{equation}
\Psi_2(\chi) = \frac{1}{6}\left[4P_2(\chi)+2P_1(\chi)\right]\;.
\label{eq:pp2_2}
\end{equation}

The general expression for $\Psi_i(\chi)$ may be obtained, iteratively,
through:

\begin{equation}
\Psi_i(\chi)=\frac{1}{i+1}\left[2\left[\int_\chi^1\,
\frac{1}{l}\Psi_{(i-1)}(l)\,dl
\right]+(i-1)\Psi_{(i-1)}(\chi)\right]\;.
\label{eq:ppi}
\end{equation}

\noindent
More especifically, if $\{a_i^{(k)}\}$ denote the coefficients of the
expansion of $\Psi_i(\chi)$ at the $i$-{\it th} step:

\begin{equation}
\Psi_i(\chi)=\sum_{k=1}^ia_i^{(k)}P_k(\chi)\;,
\label{eq:ppexp}
\end{equation}

\noindent
the above recurrence relation gives:

\begin{equation}
a_i^{(k)}=\frac{1}{i+1}
          \left[2(1-\delta_{k,1})a_{(i-1)}^{(k-1)}+(i-1)a_{(i-1)}^{(k)}
         \right]\;,
\label{eq:aexp}
\end{equation}

\noindent
where Eq.\ (\ref{eq:recnif1}) has been used and $\delta_{i,j}$ is
the usual Kronecker Delta function.
Starting from $a_1^{(1)}=1$ and $a_1^{(k)}=0$, $1 < k \le N$, one may
thus generate all the coefficients associated with the $N$-{\it th} fracture of
the system.
This procedure is extremely fast and requires a very small amount of
computational effort.

\subsection{2-dimensional models}
\label{sec:mod2d}

The 2-dimensional objects considered in this sub-section are squares of unitary
sides.
We assume that, during the fracture process, the broken pieces detach from
the body and do not fragment afterwards.
The models we discuss aim at describing some of the gross features of the
fragmentation of a square plate caused by a strong impact on one of its sides.
Therefore, straight fissures always start at the left edge of the square
and end up at one of its borders or at another crack, whichever is found first.

As in the 1-dimensional case, the number of fractures $N$ is associated with
the strength of the impact on the object.
The position of each starting point of a crack is uniformly selected along
the left edge of the object.
We consider two different pictures for the propagation of the cracks:

\begin{description}
\item{\bf a)} Non-Interacting Isotropic Fissure (NIIF2): in which
a crack grows as a straight line with equal probability along any direction
inside the square.
\item{\bf b)} Non-Interacting Anisotropic Fissure (NIAF2): where
a straight line representing a crack propagates only along selected
directions.
This version of the model aims at describing the fragmentation of
objects, such as cristaline solids, whose internal structure leads to
preferential cracking directions.
For the sake of simplicity, only two possible directions are considered,
$-\pi/4$ or $+\pi/4$, with respect to the perpendicular to the left edge of
the square.
Each of these two directions is chosen with equal probability.
\end{description}

\begin{figure}[ht]
\begin{center}
\includegraphics*[width=10cm]{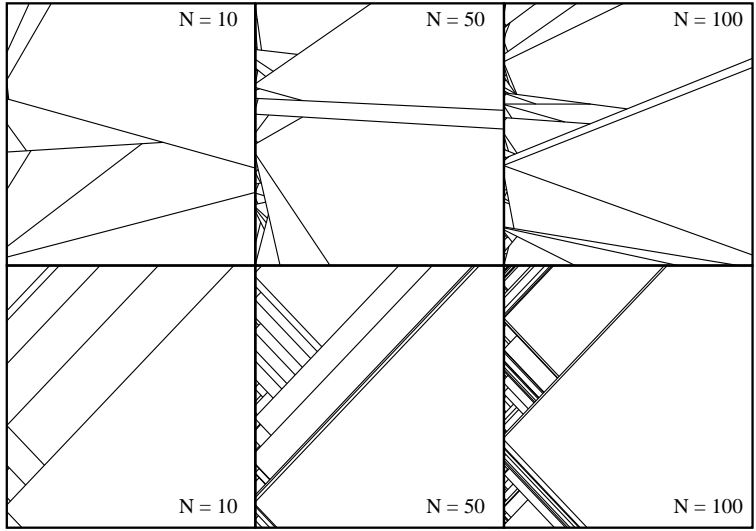}
\caption{Fragmentation patterns produced by the isotropic (upper panel)
and anisotropic (lower panel) propagation of the fractures, for $N=10$,
$N=50$, and $N=100$ cracks.}
\label{fig:fragpatterns}
\end{center}
\end{figure}

These scenarios lead to qualitatively different fragmentation patterns.
This is illustrated in Fig.\ \ref{fig:fragpatterns}, which shows
samples of the patterns obtained for different values of $N$ for the two cases
we have considered.

In contrast to the 1-dimensional cases, analytical predictions are much
more difficult in the present situation.
However, particular considerations can be made for very small areas.
Among the complex geometric objects that may be formed, triangled shaped
pieces are much more likely to contribute in this area region, at least
for $N$ not too large.
The area of the triangles is $\chi^2/2$, where $\chi$ denotes the size of
their bases.
Since the distances between two adjacent cracks $\chi$ are statistically
distributed according to Eq.\ (\ref{eq:probhom}), the probability density
$P_N(A)$ of observing a fragment of area $A$, if the system undergoes $N$
fractures, is given by:

\begin{eqnarray}
P_N(A) & = &\int_0^1\,\delta(A-\frac{\chi^2}{2})N(1-\chi)^{N-1}\,d\chi\nonumber\\
     & = &\frac{N}{\sqrt{2A}}\left[1-\sqrt{2A}\right]^{N-1}\;,\;\;A<<1\;.
\label{eq:pamsall}
\end{eqnarray}

\noindent
In this expression, we have neglected contributions from other geometric
shapes.

Similar considerations also hold for the NIAF2 model, but thin
trapezoidal shapes, such as those which appear in Fig.\ \ref{fig:fragpatterns},
also must be considered.
However, the corresponding area should also be proportional to
$\chi^2$.
Therefore, the area distribution, for small values of $A$, should also be
a power-law with exponent $-1/2$, and a similar behavior is expected to be
observed in both models.

\section{Results}
\label{sect:results}
Computational experiments have been carried out for all the models described in
the previous section.
The corresponding results are presented below, together with their
interpretations, accompanied by the analytical explanations we have obtained.

\subsection{1-dimensional models}
\label{sect:1d}
\begin{figure}[th]
\begin{center}
\includegraphics*[width=11cm]{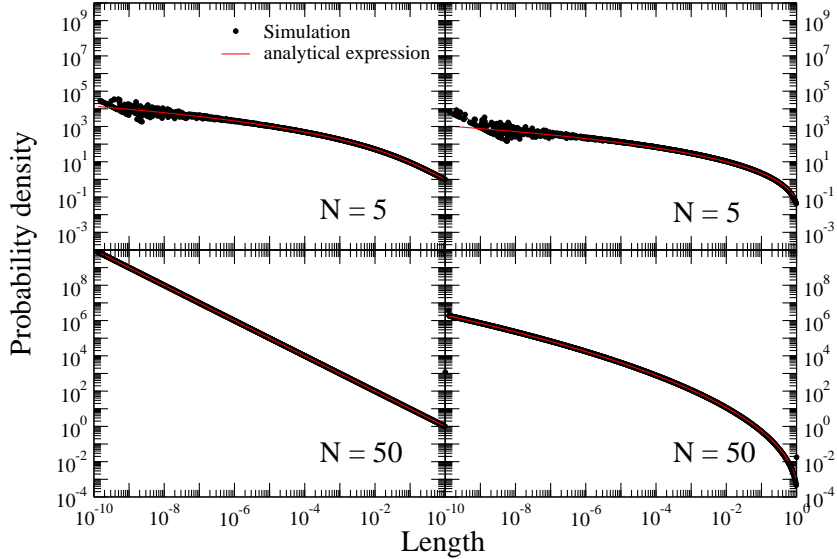}
\caption{(Color online) Simulation of 1-dimensional models compared to the
analytical predictions of Eqs.\ (\ref{eq:spronif1}) and (\ref{eq:ppi}).
Results corresponding to the  NIF1 model are shown at the left panel,
whereas those associated with the IF1 are displayed on the right
panel.}
\label{fig:comp1dm}
\end{center}
\end{figure}

The probability density of observing a fragment of size $\chi$, when the
violence of the impact on it is such that $N$ fractures are made on the system,
has been predicted in the last section, for the two 1-dimensional scenarios
we considered.
In Fig.\ \ref{fig:comp1dm}, we compare our predictions to the results obtained
in our computer simulations.

It shows that the size distributions corresponding to the different
fragmentation modes are qualitatively different.
Particularly, one notices that the interaction among the fragments leads to
smaller fragment sizes, compared to the NIF1 model.
This is quite reasonable, on physical grounds, since any further interaction,
after their formation, would reduce their average size,
compared to the case in which they do not interact.
A striking feature of the NIF1 model is that the approximation, given by
Eq.\ (\ref{eq:sumprobnif1}), holds for 10 decades, for $N$ as small as
$N=50$, as anticipated in the last section.
As a matter of fact, the analytical predictions reproduce the data
to a great degree of accuracy.

The difference between the two fragmentation scenarios may be further
illustrated by considering the average size of the
fragments generated at the $i$-{\it th} step.
In the NIF1 model, the average size of the
fragment created at the $i$-{\it th} step is given by:

\begin{equation}
\langle\chi_i\rangle = \int_0^1\,\chi P_i(\chi)\d\chi
                     = \frac{1}{2^i}\;.
\label{eq:avesizeNIF1}
\end{equation}

\noindent
To arrive at this expression, we have used Eq.\ (\ref{eq:probnif1}).

If the fragments are allowed to interact, as in the case of the
IF1 model, the average size can be estimated in the following way.
The $i$-{\it th} fissure is made with equal probability on each of the
$i$, already existent, fragments.
The average size of these fragments is equal to $1/i$.
Therefore, the size of the $(i+1)$-{\it th} fragment, created at
the $i$-{\it th} step, corresponds to:

\begin{equation}
\langle\chi_i\rangle = \frac{1}{2i}\;.
\label{eq:avesizeIF1}
\end{equation}

\begin{figure}[ht]
\begin{center}
\includegraphics*[width=11cm]{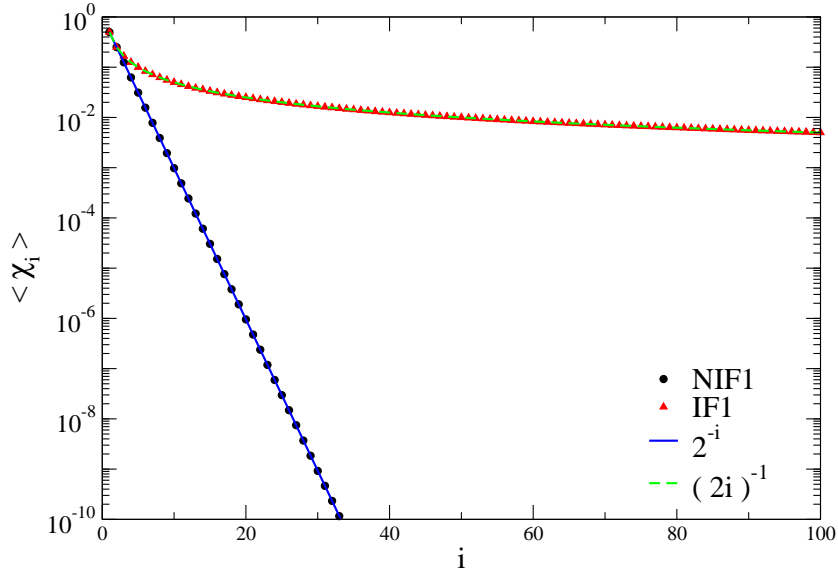}
\caption{(Color online) Comparison between the average size of the
fragments produced at the $i$-{\it th} step in the 1-dimensional models.}
\label{fig:compavesize1d}
\end{center}
\end{figure}

These results show that the average size of the fragments produced at
a given step in the IF1 model is, in general, much larger than in the case
of the NIF1 model.
This prediction is illustrated in Fig.\ \ref{fig:compavesize1d},
as well as the corresponding simulation data.
The agreement between the analytical formulae and the numerical
simulations is, once again, remarkably good.

This conclusion apparently contradicts that drawn by the analysis of the
size distributions displayed in Fig.\ \ref{fig:comp1dm}, {\it i.e.} that
the interacting fragment picture gives much smaller fragments than
in the non-interacting scenario.
Nevertheless, the apparent inconsistency disappears if one realizes that
the survival probability of the fragment produced in the NIF1 model is
1, whereas the fragments are not preserved during the breakup process in the
IF1 model.
Therefore, large fragments formed at any step in the NIF1 model remain
during the whole process, in contrast to those produced in the IF1 model.

The characteristics of the two models may be better understood by analizing
the Dalitz plot associated with their final size distributions.
This is constructed by selecting the 3 largest fragments within each event,
and calculating:

\begin{equation}
x_k=\chi_k/\sum_{k=1}^3\chi_k\;,\;\;\;\;\;\;(k=1,\, 2,\,{\rm and}\, 3)
\label{eq:Dalitz}
\end{equation}

\noindent
which represent the perpendicular distances to the $k$-{\it th} side of an
equilateral triangle.
In this equation, $\chi_k$ corresponds to the size of one of the selected
fragments.
Thus, a point associated with a given event is plotted inside the triangle
using the above expression.
To eliminate artificial structures, the indices $\{k\}$ in
Eq.\ (\ref{eq:Dalitz}) are randomized in each event.

\begin{figure}[ht]
\begin{center}
\begin{tabular}{ccc}
\includegraphics*[width=6cm]{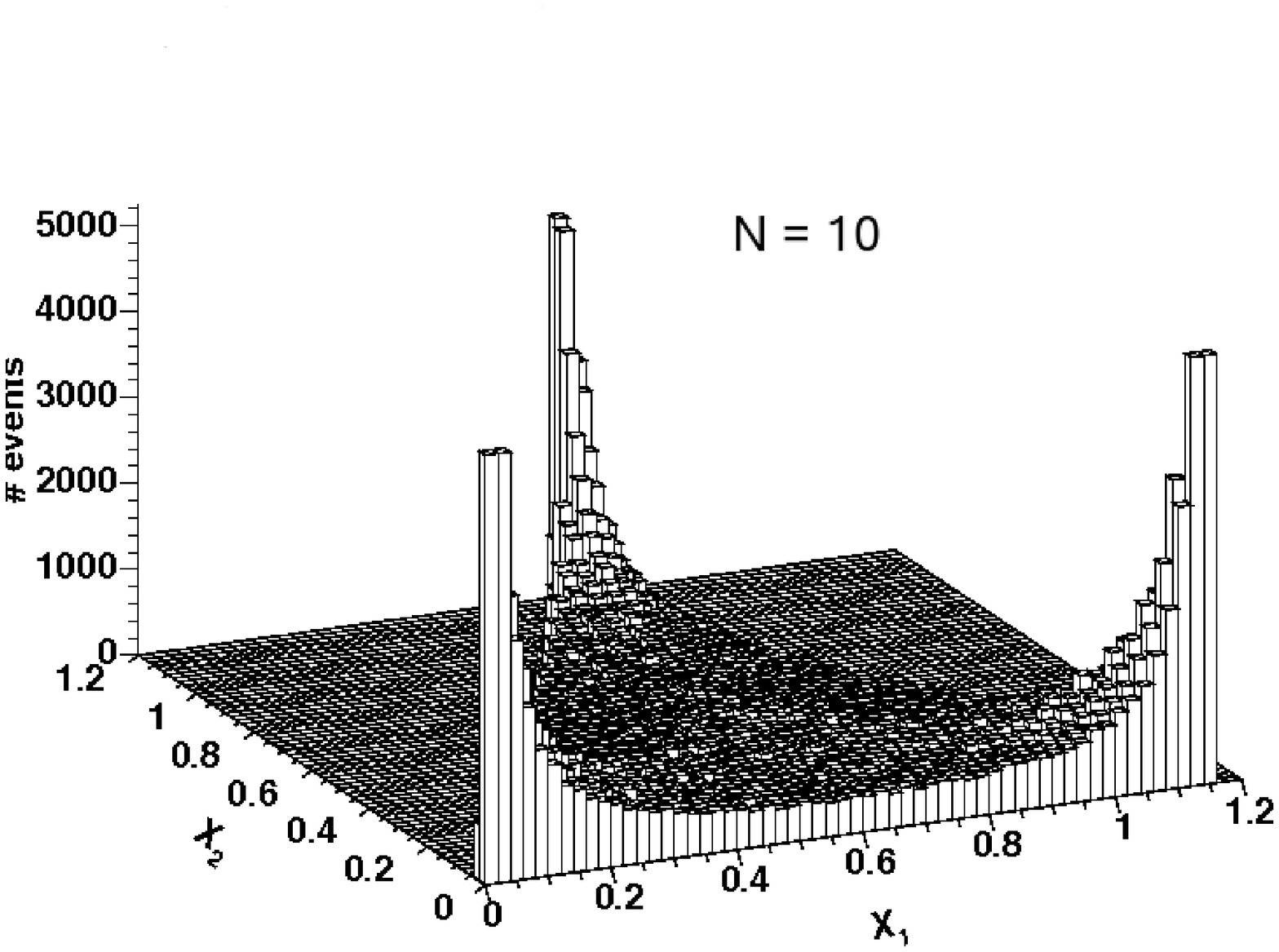} & \hskip 1.0cm &
\includegraphics*[width=6cm]{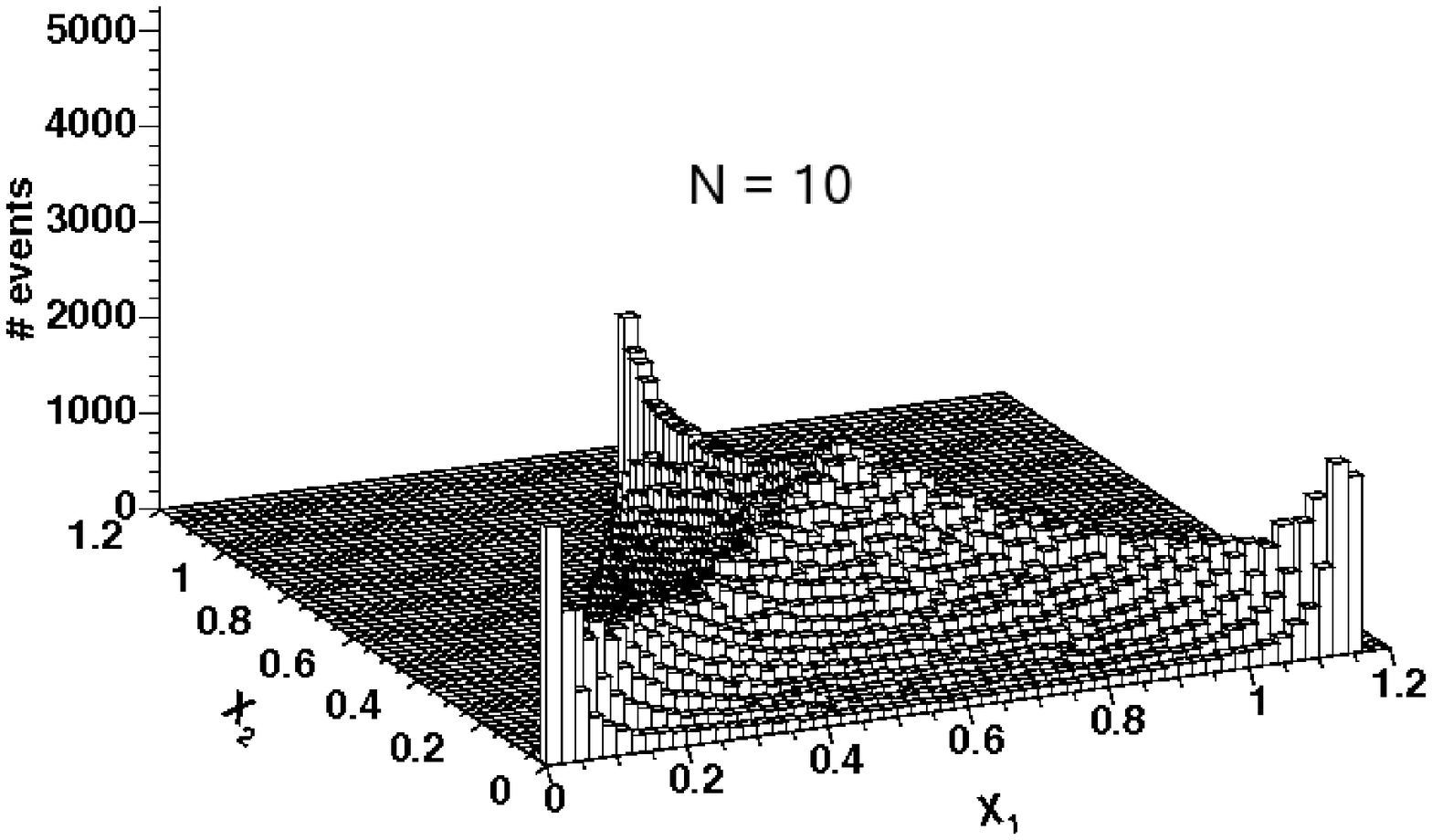} \\
\includegraphics*[width=6cm]{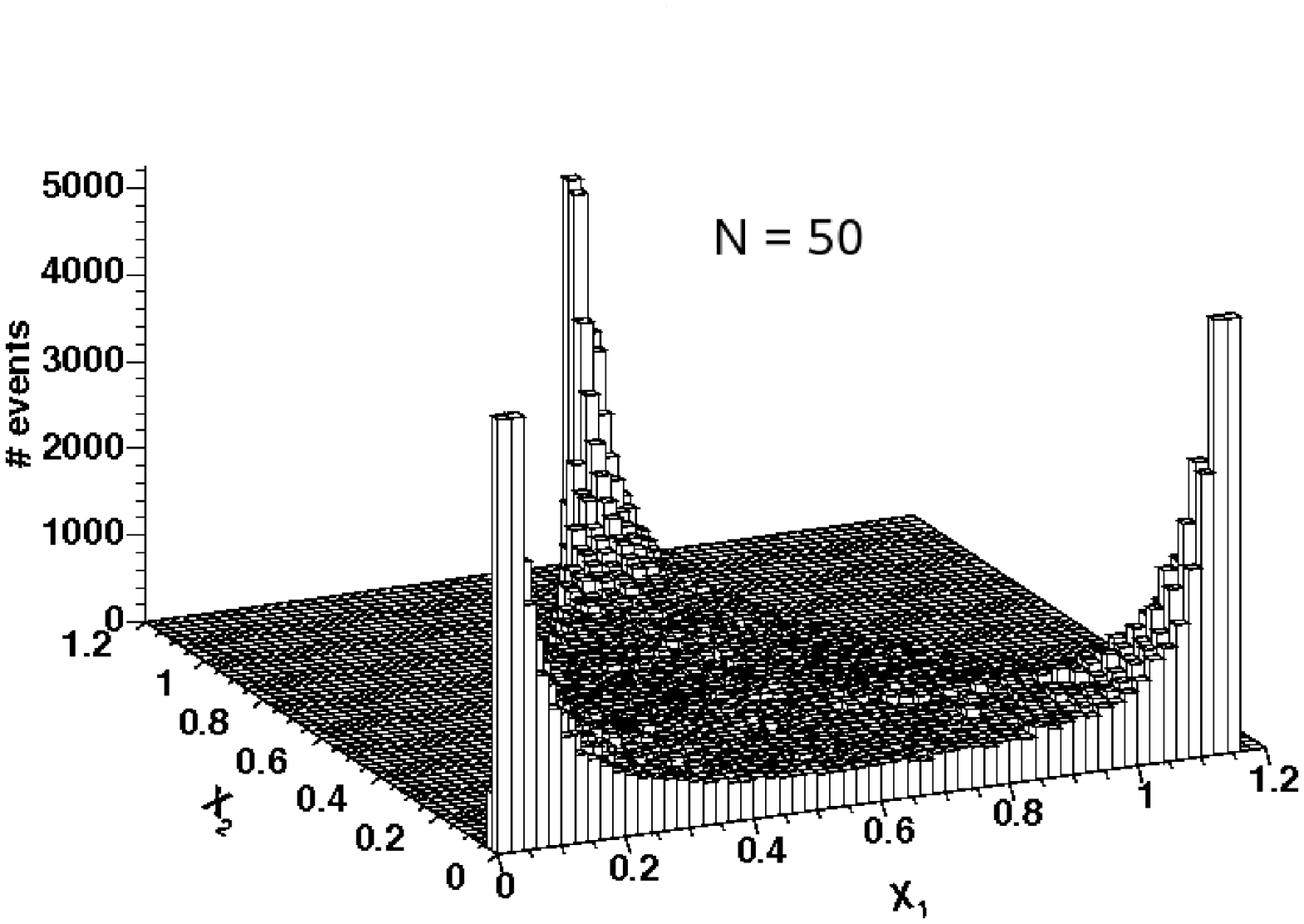} & &
\includegraphics*[width=6cm]{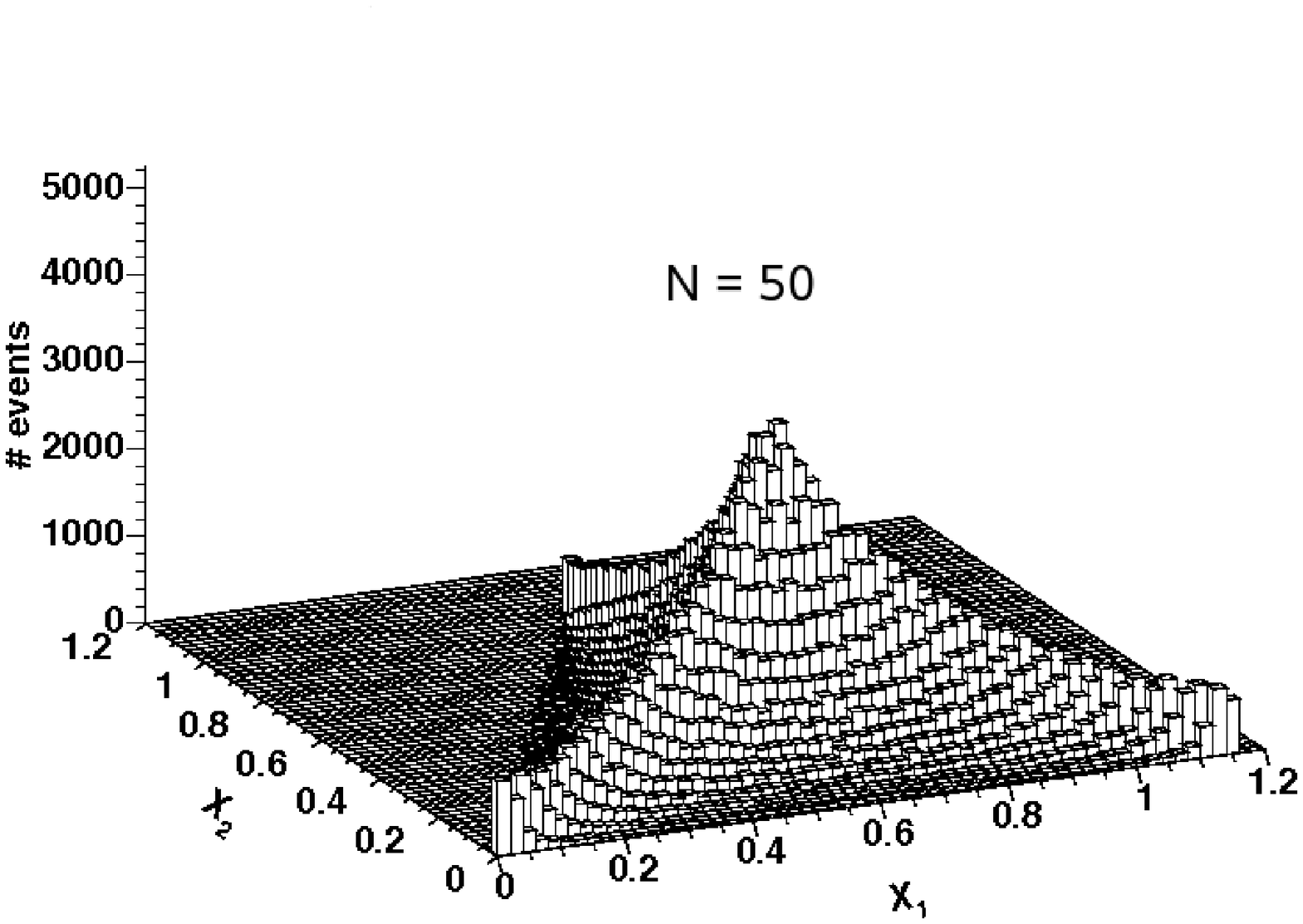} \\
\includegraphics*[width=6cm]{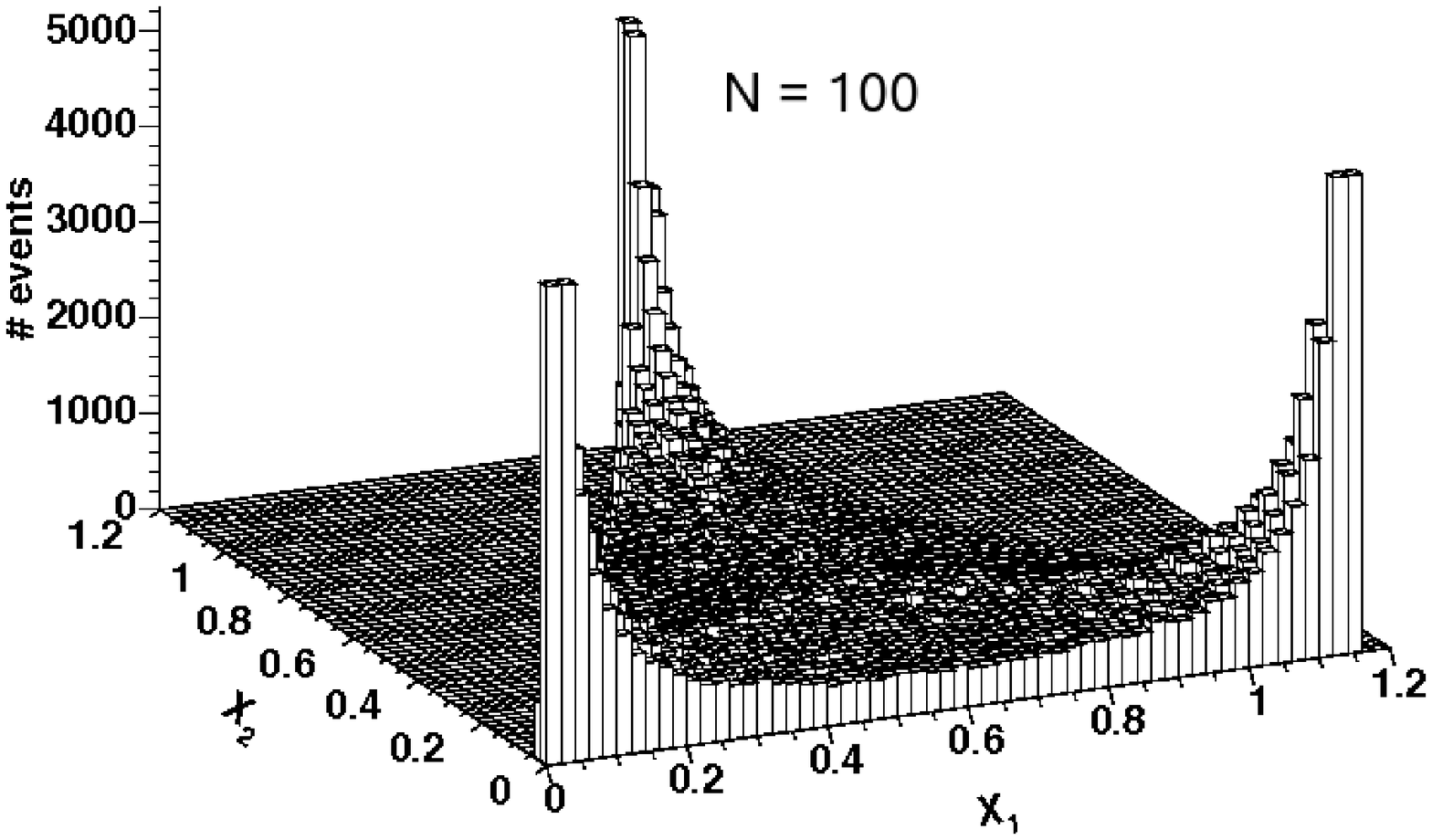} & &
\includegraphics*[width=6cm]{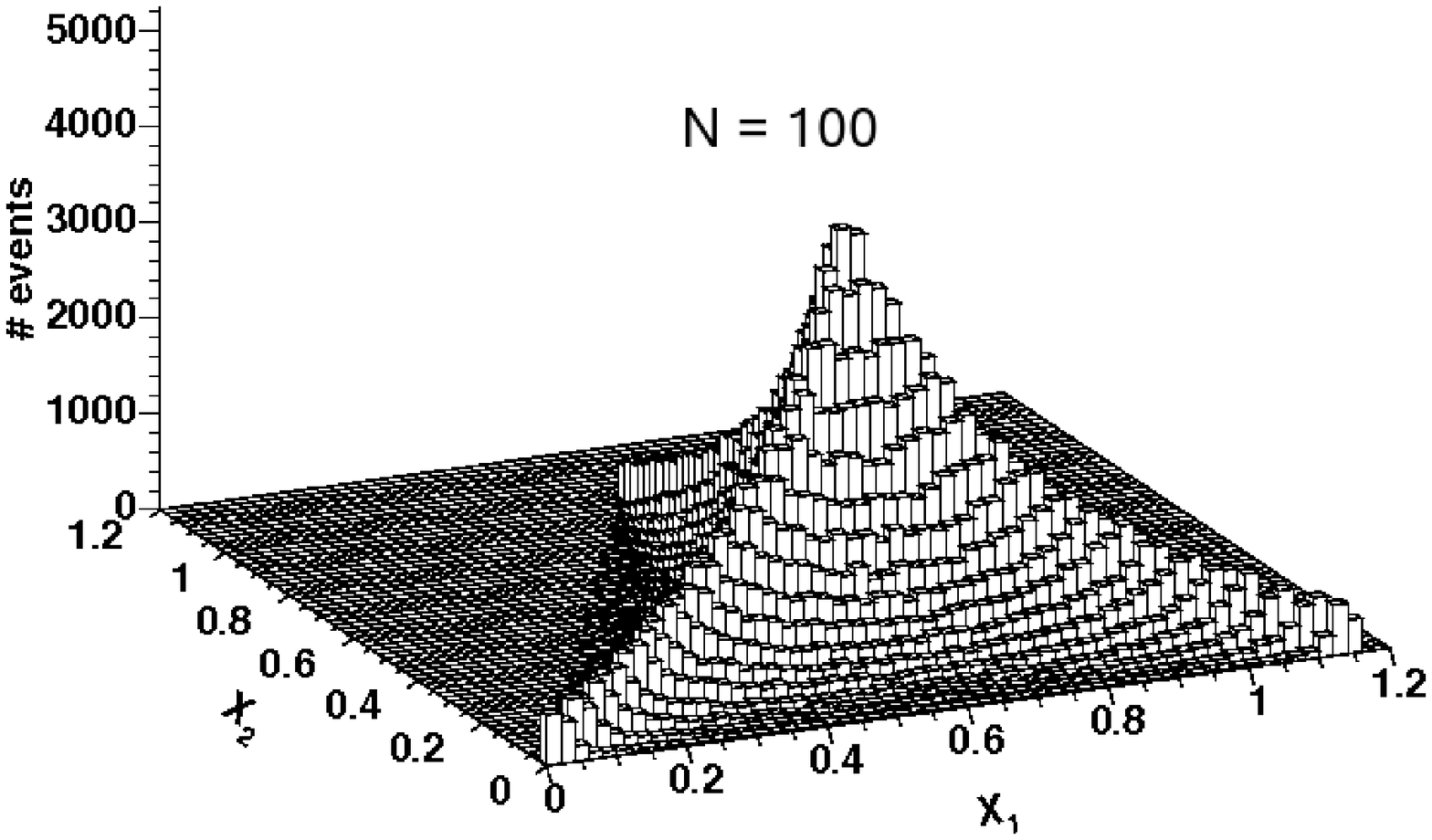} 
\end{tabular}
\caption{Dalitz plots associated with the size distributions of
the NIF1 (left panels) and IF1 (right panels) models.
For details, see text.}
\label{fig:Dalitz}
\end{center}
\end{figure}

By construction, all the points must lie inside the triangle.
When the size distribution is dominated by a large fragment, whereas the
others are much smaller, one should observe bumps close to the vertices.
On the other hand, if the 3 largest fragments have approximately the
same size, the peak of the distribution is found near the center of
the triangle.
Finally, if the size of two of the selected fragments is similar, but
much larger than that of the third one, bumps close to the middle point
of the triangle sides should be observed.

The results shown in Fig.\ \ref{fig:Dalitz} reveal that a large fragment
is always surounded by smaller ones in the NIF1 model, in contrast to
the IF1 model in which the three largest fragments have approximately
the same size.
Thus, the two fragmentation pictures lead to qualitatively different
size distributions.

\subsection{2-dimensional models}
\label{sect:2d}
\begin{figure}[ht]
\begin{center}
\includegraphics*[width=11cm]{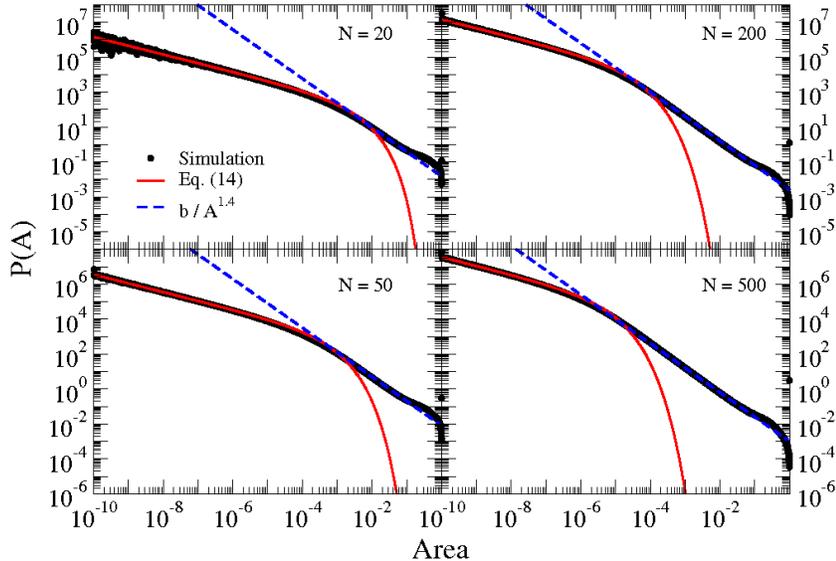} 
\caption{(Color online) Probability density of observing a fragment of area $A$,
obtained with the NIIF2 model for different values of $N$.
For details, see text.}
\label{fig:areaDC}
\end{center}
\end{figure}

Computer simulations have also been carried out for the fragmentation of 
square objects, described in sect. \ref{sec:mod2d}.
The results associated with the NIIF2 model are depicted in
Fig.\ \ref{fig:areaDC}, for different values of $N$.
It is clear that the area distribution exhibits two power-law regimes.
One of them has been already predicted in the last section, corresponding
to $A^{-1/2}$, and is represented by the full lines in this figure.
The agreement with the analytical prediction is very good in all cases.
However, one notices that deviations appear as $N$ increases.
This behavior should be expectedi since, as already mentioned, when more
cracks are made on the system, the contribution of objects more complex
than triangles becomes non-negligible for small areas.

One may also notice that another power-law regime, of expoment
approximately equal to $-1.4$, appears after the breakdown of
Eq.\ (\ref{eq:pamsall}), in the area regions where no particular
geometric shape should dominate the distribution.
This is illustrated by the dashed lines in this figure.
Since we could not obtain any analytical explanation for this power-law
behavior, this is an empirical finding.
Nevertheless, it agrees qualitatively with the fractal analysis made
in ref.\ \cite{fractalKim1995} as well as with experimental studies
\cite{FragBohr1993} on the fragmentation of rectangular objects, which have
demonstrated that the fragment
mass distribution (of relatively large pieces) obeys a power-law with exponent
approximately equal to 1.15 -- 1.20.
However, those experiments focused on impacts equally distributed over
the surface of the plate.
Since our model is intended to describe a different picture,
the small discrepancy between our results and the experimental
observations is quite reasonable.

The qualitative difference between our results and those just mentioned
is the existence of two power-law regimes in our model, whereas only one
has been observed in the experiments.
Unfortunately, it is rather difficult to investigate our prediction
experimentally since the second power-law regime should appear
for very small values of the area.

\begin{figure}[ht]
\begin{center}
\includegraphics*[width=11cm]{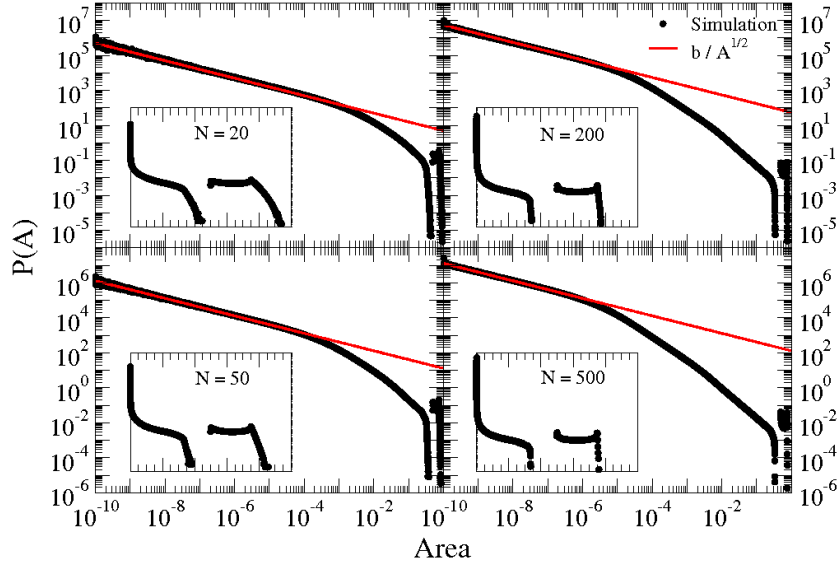} 
\caption{(Color online) Probability density of observing a fragment of area
$A$, obtained with the NIAF2 model for different values of $N$.
The inserts correspond to log-linear plots of the respective area
distributions.
For details, see text.}
\label{fig:areaDDin}
\end{center}
\end{figure}

The behavior observed in the anisotropic model (NIAF2) is qualitatetively
different from the former case, as is illustrated in Fig.\ \ref{fig:areaDDin}.
The power-law regime, with exponent $-1/2$, predicted in the last section
is indeed observed, but the area distribution tends to develop a
gap for $1/3 < A < 1/2$, whereas the production of fragments whose
$A>0.75$ is hindered.
This trend becomes more and more pronounced as $N$ increases.

\begin{figure}[ht]
\begin{center}
\includegraphics*[width=10cm]{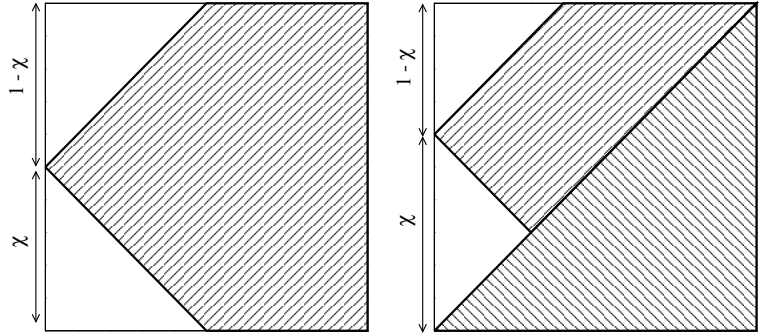}
\caption{Fragments of large areas that could be observed in the
end of the process for $N$ large.}
\label{fig:survivingAreas}
\end{center}
\end{figure}

This feature of the area distribution has a simple explanation, based on
the geometric constraints imposed by the anisotropy.
In the limit of large $N$, only a few particular shapes may contribute
to large areas.
More specifically, only those corresponding to the hatched
areas in Fig.\ \ref{fig:survivingAreas} would not undergo further
fractures and survive until the end of the process.

\noindent
The area corresponding to the pentagon shown in the left panel of this
figure is:

\begin{equation}
A(\chi)=1-\frac{\chi^2}{2}-\frac{(1-\chi)^2}{2}\;,
\label{eq:areaTrap1}
\end{equation}

\noindent
whereas that associated with the trapezoid displayed in the right panel
reads:

\begin{equation}
A(\chi)=\frac{1}{2}-\frac{\chi^2}{4}-\frac{(1-\chi)^2}{2}\;.
\label{eq:areaTrap2}
\end{equation}

\noindent
The area of the former lies between $1/2 \le A \le 3/4$, whereas for
the latter one has $0 \le A \le 1/3$.
The only big triangle which would not be cut into pieces for large $N$ is
the one shown in the right panel, but contributes to the area distribution
at the right hand side of the gap.
Other shapes have vanishing small survival probability, for large $N$, and
would, therefore, be broken into pieces, contributing only to small
areas.
Thus, the preferred cracking directions give rise to the peculiarities observed
in the area distributions.
We have checked that qualitatively similar results are also obtained for
other cracking directions.

\begin{figure}[ht]
\begin{center}
\includegraphics*[width=11cm]{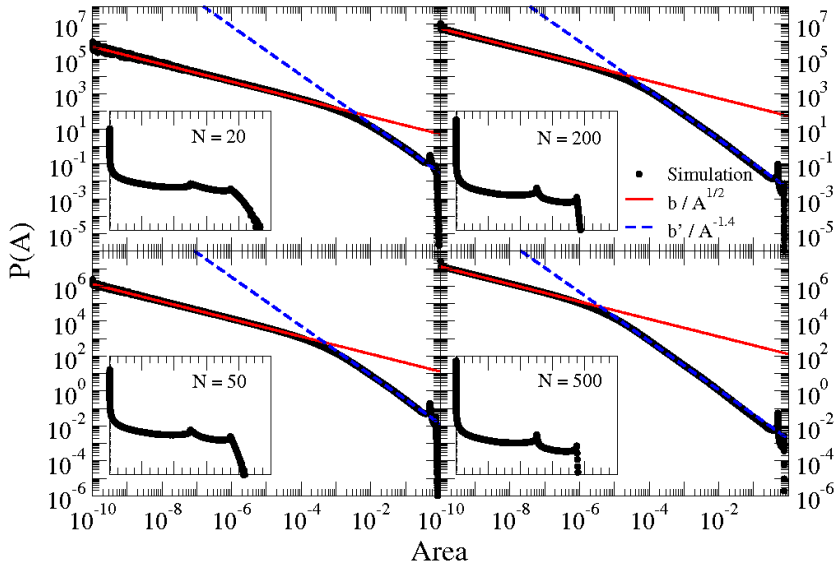} 
\caption{(Color online) Probability density of observing a fragment of
area $A$, obtained with the NIAF2 model for different
values of $N$, in the case that 3 cracking directions are allowed.
The inserts correspond to log-linear plots of the respective area
distributions.
For details, see text.}
\label{fig:areaD3A}
\end{center}
\end{figure}

In a more realistic model, fractures should develop preferentially along the
selected directions but, with lower probability, other directions should
also be allowed.
However, one still finds fingerprints on the area distributions.
This is illustrated in Fig.\ \ref{fig:areaD3A}, where the area distributions
for different values of $N$ are shown, in the case where one allows 3
equiprobable cracking directions:
$\theta=-\frac{\pi}{4}$, $\theta=0$, and $\theta=+\frac{\pi}{4}$.
Although the gap disappears, due to fissures corresponding to $\theta=0$,
the first derivative of $P(A)$ shows a singularity at $A=1/2$.
One should also notice that, for large areas, but $A<1/2$, the distribution
follows the same power-law found in the isotropic case.

\section{Concluding remarks}
\label{sect:conclusions}
The schematic models developed in this work for the fragmenation of
one dimensional objects revealed that qualitatively different size
distributions should be observed whether the fragments interact or
not during the breakup process.
Size distributions obeying a power-law are only obtained in the
non-interacting scenario.
Intuitively, this finding is quite reasonable since the interaction among the
fragments should lead to the appearance of a scale.

The area distributions, for not too small areas, predicted by our isotropic
models are in reasonable agreement with those predicted in
\cite{fractalKim1995} and obtained experimentally in \cite{FragBohr1993}.
However, from simple considerations, we predict that the probability
density of finding fragments of small areas should be a power-law
with exponent 1/2, in both isotropic and anisotropic fragmenting pictures. 
Owing to the difficulties in measuring small fragments, we are not
aware of experimental observations in this area region, which could be
confronted with our predictions.

Our results also suggest that clear signatures of anisotropy effects
should be found in the area distributions, for not too small values of $A$.
Therefore, we suggest that the fragmentation of anisotropic materials
be studied to provide more insight in this phenomenon.

\begin{flushleft}
\bf
Acknowledgements
\end{flushleft}
We would like to thank Dr.\ K.\ Sneppen, Dr.\ P.M.C. de Oliveira,
and Dr.\ C.E. Aguiar for many fruitful discussions.
We would like to acknowledge CNPq, FAPERJ, and
PRONEX(CNPq-FAPERJ) under contract 26.171.176.2003, for
partial finantial support.

\bibliography{artigo}

\end{document}